\documentclass[11pt]{article}

\usepackage{amsmath,amssymb,amsthm,verbatim,fullpage}
\usepackage{epsfig,color}

\theoremstyle{plain}
\newtheorem{theorem}{Theorem}
\newtheorem{lemma}{Lemma}
\newtheorem{corollary}[lemma]{Corollary}

\newtheorem{proposition}[lemma]{Proposition}

\theoremstyle{definition}
\newtheorem{definition}{Definition}

\newcommand{\E}{\mathbb{E}}
\newcommand{\la}{\langle}
\newcommand{\ra} {\rangle}
\newcommand{\eps}{\varepsilon}

\newcommand{\algon}{\ensuremath{\mathcal{A}}}
\newcommand{\A}{\ensuremath{\mathcal{A}}}
\newcommand{\opt}{\ensuremath{\text{\textsc{Opt}}}}
\newcommand{\cost}{\text{cost}}
\newcommand{\bi}{\begin{itemize}}
\newcommand{\ei}{\end{itemize}}
\newcommand{\be}{\begin{enumerate}}
\newcommand{\ee}{\end{enumerate}}
\newcommand{\ie}{\textsl{i.e.}}
\newcommand{\Ie}{\textsl{I.e.}}

\newcommand{\poly}{\rm poly}
\newcommand{\f}{\mathsf{f}}

\newcommand{\mommit}[1]{}

\DeclareMathOperator{\lca}{lca}

\title{Multi-Embedding of Metric Spaces%
\thanks{A preliminary version of this paper appeared in \cite{BM03}.}}

\author{Yair Bartal\thanks{School of Computer Science, Hebrew University, 
Jerusalem 91904, Israel.
email: \texttt{yair@cs.huji.ac.il}.
{Supported in part by a grant from the Israeli Science Foundation 
(195/02).}}
\and
Manor Mendel\thanks{School of Computer Science, Hebrew University, Jerusalem 
91904, Israel.
Email: \texttt{mendelma@gmail.com}.
{Supported by the Landau Center and a grant from the Israeli Science 
Foundation
(195/02).}}}

\begin{document}

\maketitle

\begin{abstract}
Metric embedding has become a common technique  in the design of
algorithms. Its applicability is often dependent on how high the
embedding's distortion is. For example, embedding finite metric space into
trees may require linear distortion as a function of its size. Using
probabilistic metric embeddings, the bound on the distortion
reduces to logarithmic in the size.

We make a step in the direction of bypassing the lower bound on
the distortion in terms of the size of the metric. We define
``multi-embeddings" of metric spaces in which a point is mapped
onto a set of points, while keeping the target metric of
polynomial size and preserving the distortion of paths. The
distortion obtained with such multi-embeddings into ultrametrics
is at most $O(\log \Delta\log\log \Delta)$ where $\Delta$ is the
\emph{aspect ratio} of the metric. In particular, for expander
graphs, we are able to obtain {\em constant} distortion embeddings
into trees in contrast with the $\Omega(\log n)$ lower bound for
all previous notions of embeddings.

We demonstrate the algorithmic application of the new embeddings
for two optimization problems: \emph{group Steiner tree} and \emph{metrical 
task
systems}.
\end{abstract}

\begin{comment}
\begin{keywords}
Metric embeddings, Group Steiner tree, Metrical task systems
\end{keywords}

\begin{AMS}
68W25, 68R10
\end{AMS}

\pagestyle{myheadings}
\thispagestyle{plain}
\markboth{Y. BARTAL AND M. MENDEL}{MULTI-EMBEDDING OF METRIC SPACES}
\end{comment}

%\clearpage

\section{Introduction}

Finite metric spaces and their analysis play a significant role in
the design of combinatorial algorithms. Many algorithmic
techniques were introduced in recent years concerning and using
metric spaces and their approximate embedding in other spaces, see
the surveys \cite{Ind01,IM04} for an overview of this topic.

%A metric space $M$ is a pair $(S,d)$, where $S$ is a set of points and
%$d:S\times S \rightarrow \mathbb{R}^+$ is a distance function that 
% satisfies:
%(i) $d(u,v)\geq 0$ for $u,v \in S$; (ii) $d(u,v)=0$ if and only if $u=v$;
%(iii) $d(u,v)=d(v,u)$ for all $u,v\in S$; (iv) $d(u,v)+d(v,w)\geq d(u,w)$ 
% for
%all $u,v,w\in S$. $d$ is also called the metric of $M$.

\begin{definition}
\label{def:embedding}
An embedding of a metric space $M=(V_M,d_M)$
into a metric space $N=(V_N,d_N)$ is a mapping $\phi:V_M
\rightarrow V_N$. The embedding is called \emph{non-contractive}
if for all $u,v \in V_M$, $d_M(u,v) \leq d_N(\phi(u),\phi(v))$ and
has distortion at most $\alpha$ if in addition for all $u,v \in
V_M$, $d_N(\phi(u),\phi(v)) \leq \alpha \cdot d_M(u,v)$. A
non-contractive embedding whose distortion is at most $\alpha$ is
called $\alpha$-embedding.
\end{definition}

The general framework for
applying metric embeddings in optimization problems is to embed a
given metric spaces into a metric space from some ``nice" family
and then apply an algorithm for that space. As a result, the
approximation ratio increases by a factor equal to the embedding's
distortion.

Among others, embeddings into low dimensional normed spaces
\cite{Bou85,LLR95} as well as probabilistic embeddings into trees
\cite{Bar96,Bar98,FRT03,Bar03} have many algorithmic applications.
In both cases the distortions of the embeddings are logarithmic in
the size of the metric. Unfortunately, there is a matching lower
bound on the distortion of these embeddings as well, which sets a
limit to their applicability. This paper presents a partial remedy
for this problem.

Tree metrics, and in particular {ultrametrics}, seem a natural
choice as a target class of ``simple" metric spaces.
Unfortunately, standard embedding is not useful when the target
space is a tree metric. Embedding arbitrary metric spaces into
trees requires distortion linear in the size of the metric
space~\cite{RR98}. Probabilistic embedding~\cite{Bar96} provides a
way to bypass this problem:

\begin{definition}[Probabilistic Embeddings] \label{def:prob-approx}
A metric space $M=(V_M,d_M)$ is $\alpha$-\emph{probabilistically
embedded} in a set of metric spaces $\mathcal{S}$ if there exists
a distribution $\mathcal{D}$ over $\mathcal{S}$ and for every
$N\in\mathcal{S}$, a non-contractive embedding $\phi_N:V_M
\rightarrow V_N$, such that for all $u,v \in V_M$, $\E_{N\in
\mathcal{D}}[d_N(\phi_N(u),\phi_N(v))] \leq \alpha \cdot
d_M(u,v)$.
\end{definition}

Using probabilistic embeddings, it is possible to obtain much
better bounds on the distortion
\cite{AKPW95,Bar96,Bar98,FRT03,Bar03}. The following bound is
shown in \cite{FRT03,Bar03}:

\begin{theorem}  \label{thm:bar98}
Any metric space on $n$ points can be  $O(\log n )$
probabilistically embedded in a set of $n$-point {ultrametrics}.
Moreover, the distribution can be sampled efficiently.
\end{theorem}

Theorem~\ref{thm:bar98} found many algorithmic applications in approximation
algorithms, online algorithms, and distributed algorithms, see for
example~\cite{Bar96,GKR98,FM00,KT99,BCR01}. The bound on the distortion in
Theorem~\ref{thm:bar98} is tight even for probabilistic embeddings into tree
metrics for which there is an $\Omega(\log n)$ lower bound \cite{Bar96}.

Theorem~\ref{thm:bar98} was originally formulated for a class of metric
spaces defined by the following natural generalization of ultrametrics:

\begin{definition}[\cite{Bar96}] \label{def:hst}
For $k\geq 1$, a $k$-\emph{hierarchically well-separated tree}
($k$-HST) is a metric space defined on the leaves of a rooted tree
$T$. To each vertex $u\in T$ there is associated a label
$\Delta(u) \ge 0$ such that $\Delta(u)=0$ if and only if $u$ is a leaf of
$T$. The labels are such that if a vertex $v$ is a child of a
vertex $u$ then $\Delta(v)\leq \Delta(u)/k$. The distance between
two leaves $x,y\in T$ is defined as $\Delta(\lca(x,y))$, where
$\lca(x,y)$ is the least common ancestor of $x$ and $y$ in $T$.
\end{definition}

The definition of finite ultrametric is the same as a 1-HST. Any $k$-HST
is therefore, in particular, an ultrametric and any finite ultrametric
can be $k$-embedded in some $k$-HST \cite{Bar98}. We can
therefore restrict our attention to ultrametrics, while all
results generalize to $k$-HSTs.

The main contribution of this paper is in offering a new type of metric
embedding that makes it possible to bypass lower bounds for the standard and
even probabilistic metric embeddings. There are two key observations that
lead to this new type of embedding. The first is that in some applications 
it
is natural to match a point onto a set of points in the target metric space.
Motivated by two applications of Theorem~\ref{thm:bar98}: the
\emph{group Steiner tree problem} (henceforth, GST), and the
\emph{metrical task systems problem} (henceforth, MTS),
we propose the following definition:

\begin{definition}[Multi Embedding] \label{def:multi-embedding}
A multi embedding of $M$ in $N$ is a partial surjective function
$\f$ from $N$ on $M$, i.e. each point $x\in M$ is embedded into a
non-empty set $\f^{-1}(x)$. Points in $\f^{-1}(x)$ are called
\emph{representatives} of $x$ in $N$.
\end{definition}

The role of $\f^{-1}$ in Definition~\ref{def:multi-embedding} is
analogous to the role of $\phi$ in the
Definitions~\ref{def:embedding} and \ref{def:prob-approx} of
embedding and probabilistic embedding. Another way to define multi
embedding is by $\phi:M \to 2^N$, in which $\phi(u)\cap
\phi(v)=\emptyset$ for every $u\neq v$. In our notation we have
$\phi(x) = \f^{-1}(x)$. Since the $\f$ notation will be more
convenient, henceforth we will exclusively use it.

The second observation is that for many applications, including
those mentioned above, there is no need to approximate the
original distance for every pair of representatives. What is
really needed is that every path in the original space will be
approximated well by \emph{some} path in the target space.

A path in a metric space is an arbitrary finite sequence of points in the
space. The length of a simple path $p=\la u_1,u_2,\ldots,u_m \ra$ in a 
metric space
$M=(V,d)$ is defined as \( \ell(p)=\sum_{i=1}^{m-1} d(u_i,u_{i+1}) \).

\begin{definition}[Path Distortion] \label{def:path-approx}
A multi-embedding $\f$ of $M$ in $N$, is called \emph{non-contractive}
if for any $u,v\in N$, $d_N(u,v) \geq d_M(\f(u),\f(v))$.
The path-distortion of a non-contractive multi-embedding of $M$ in $N$, 
$\f:N\to M$,  is
the infimum over $\alpha$, for which
any path $p=\la
u_1,u_2,\ldots,u_m \ra$ in $M$, has a path $p'=\la
u'_1,u'_2,\ldots,u'_m \ra $ in $N$ such that $\f(u'_i)=u_i$ and
$\ell(p') \leq \alpha \cdot \ell(p)$.

A multi embedding  whose path-distortion is
at most $\alpha$ is called $\alpha$-path embedding.
\end{definition}

A crucial parameter for
multi-embeddings is the size of the target space
$\Gamma$.
In general, it will be desirable that
$\Gamma$ will be polynomial in the size of the source space.
In fact, if $\Gamma = \infty$ then there is a simple
$1$ path embedding of any finite metric space by trees:
Take all finite paths, convert each path to a simple path (by duplicating
points, if necessary), and put them under a single root with an edge of
length half the diameter.
This motivates a study of
the trade-off between $\Gamma$ and the path distortion of arbitrary metric
spaces by tree metrics. In Section~\ref{sec:trees} we study path
embedding of expander graphs and the hypercube into tree metrics. We show
e.g. that an $n$-point Ramanujan graphs have 3-path-embedding
into tree metrics of size
$\Gamma(n)\leq\text{poly}(n)$. This is in sharp contrast to the
status of expander graphs for previous notions of embeddings for which they
are considered ``worst case" examples with $\Omega(\log n)$ distortion
\cite{LLR95}. These results directly imply nearly tight results on the
approximation ratio for GST on expander graphs and hypercubes.

We consider multi embeddings when the class of target metric
spaces are {ultrametrics}. First, we observe that probabilistic
embedding into {ultrametrics} directly implies a bound for path
embedding by putting all the trees in $\mathcal{S}$ (the set used
in the probabilistic embedding) under a common new root. This
results with an $\alpha$-path embedding into {an ultrametric} of
size $ |\mathcal{S}|n$. Using the bound of \cite{CCGGP98} on the
number of {ultrametrics} needed in Theorem~\ref{thm:bar98} we
obtain an $O(\log n )$ path embedding into {an
ultrametric} of size $O(n^2 \log n)$.%
\footnote{In a preliminary version of this paper~\cite{BM03}, we
also introduced the notion of \emph{probabilistic
multi-embedding}. Using that notion we were able to show
probabilistic multi-embedding into {ultrametrics} of polynomial
size and path distortion $O(\log n \log\log \log n)$. Since an
$O(\log n)$ bound now follows from
Theorem~\ref{thm:bar98}~\cite{FRT03,Bar03}, we have decided to
drop the probabilistic multi-embedding result from this version.}

An important parameter of the metric spaces appearing in practice
is the aspect ratio of the metric, which is the ratio between the
diameter and minimum non-zero distance in the metric space. It
will be convenient for us to assume that the minimum distance is
1, and so the aspect ratio becomes the diameter. It turns out that
the aspect ratio of the metric plays a significant role in the
path distortion of multi-embeddings. In
Section~\ref{sec:path-approx} we prove:

\begin{theorem} \label{thm:path-approx}
Fix $\beta>1$. For any metric space $M=(V,d)$ on $|V|=n$ points
and aspect ratio $\Delta$, there exists an efficiently
constructible multi-embedding into an ultrametric of size
$n^\beta$, whose path distortion is at most
\[
O_\beta( \min\{\log n \cdot \log \log n, \;\log \Delta \cdot \log\log \Delta 
\}). \]
\end{theorem}

Our construction beats the probabilistic embedding based
constructions on metrics with small aspect ratio. Expander graphs
are examples where a lower bound of $\Omega(\Delta)$ exists on
probabilistic embedding using trees \cite{LLR95}.

The constructions of multi-embeddings are in a sense dual to
Ramsey-type theorems for metric spaces \cite{BBM01,BLMN03}, where
the goal is to find a large subset which is well approximated by some
{ultrametric}.
% , and some of the techniques in the proof of
% Theorem~\ref{thm:path-approx} are dual to those of
% \cite{BBM01,BLMN03}.

We also provide a simple example in which
Theorem~\ref{thm:path-approx} is almost tight: Any $\alpha$-path
embedding into {ultrametrics} of a simple unweighted path of
length $n$ has $\alpha=\Omega(\log n)$. It follows, in particular,
that any $\alpha$-path embedding into {ultrametrics} of the metric
defined by an unweighted graph of diameter $\Delta$ has
$\alpha=\Omega(\log \Delta)$. 
Path embedding is motivated by two
intensively-studied algorithmic minimization problems: GST and
MTS, mentioned above. For both, the best known algorithms use
probabilistic embedding into trees/ultametrics. In
Section~\ref{sec:appl} we prove that in order to reduce these
problems to other metric spaces it is sufficient to use path
embedding. We therefore achieve improved algorithms for these
problems whenever the path embedding distortion beats that of
probabilistic embedding, and in particular, when the underlying
metric is of small aspect ratio.

\section{Applications} \label{sec:appl}

In this section we define MTS and GST
show that path distortion of multi-embeddings reduces
these problems to similar problems with different underlying metrics.

Metrical Task Systems (MTS)~\cite{BLS92} was introduced as a
framework for many online minimization problems. A MTS on metric
space $M=(S,d)$, $|S|=n$, is defined as follows. A ``system" has a
set of $n$ possible internal states $S$.  It receives a sequence
of \emph{tasks} $\sigma=\tau_1\tau_2\cdots \tau_m$. Each task
$\tau$ is a vector $\tau:S\rightarrow \mathbb{R}^+ \cup \{
\infty\}$ of nonnegative costs for serving $\tau$ in each of the
internal states of the system. The system may switch states (say
from $u$ to $v$), paying a cost equal to the distance $d(u,v)$ in
$M$, and then pays the service cost $\tau(v)$ associated with the
new state. The major limiting factor for the system is the
requirement to process the sequence in an online fashion, \ie,
serving each task without knowing the future tasks.

As customary in the analysis of online algorithms, MTS
is analyzed using the notion of \emph{competitive ratio}. A randomized 
online
algorithm $A$ is called $r$-competitive if there exists some constant $c$ 
such that for any
task sequence $\sigma$, $\E[\cost_A(\sigma)]\leq r\cdot 
\cost_{\opt}(\sigma)+c$,
where $\cost_A(\sigma)$ is the random variable of the cost for serving
$\sigma$ by $A$, and $\cost_{\opt}(\sigma)$ is the optimal (offline) cost 
for
serving $\sigma$. The current best online algorithm for the MTS problem in
$n$-point metric spaces is $O(\log^2 n \log \log n)$ competitive
\cite{FM00,FRT03} (an improvement of \cite{BBBT97,Bar98}). Both papers
\cite{BBBT97,FM00} actually solve the MTS problem for {ultrametrics}, and
then reduce arbitrary metric spaces to {ultrametrics} using
Theorem~\ref{thm:bar98}. We next show that path embedding suffices:

\begin{proposition} \label{prop:MTS-appl}
Assume that a metric space $M$ is $\alpha$-path
embedded in $N$.
Assume also that $N$
has an $r$-competitive MTS algorithm. Then
there is an $\alpha r$-competitive algorithm
for $M$.
\end{proposition}
\begin{proof}
We construct an online algorithm $\algon$ for $M$ as
follows:
Let $\algon_{N}$ be an
$r$-competitive online algorithm for $N$, and $\f:N\to M$ an $\alpha$ path 
embedding of $M$ in $N$.
The task
sequence $\sigma$ is translated to a task sequence $\sigma^N$ for $N$ task 
by
task as follows.
A task $\tau$ for $M$ is translated into a task $\tau^N$ for
$N$ such that $\tau^N(u')=\tau(\f(u'))$. $\algon$ maintains the
invariant that if $\algon_{N}$ is in state $v'$, then $\algon$ is in
state $\f(v')$.

It is easy to verify that $\cost_{\algon}(\sigma)\leq
\cost_{\algon_{N}}(\sigma^N)$, since the service costs are the same,
and the distances in $N$ are larger. Consider $\opt(\sigma)$, it
defines a path $p$ of serving $\sigma$ in $M$. Thus there exists a path 
$p^N$ as
in the statement of Definition~\ref{def:path-approx}. The path
$p^N$ is the way $\sigma^N$ would be served in $N$. In this way, since
$\f(p^N)=p$, the service costs in $N$ are the same as the services costs
in $M$, and $\ell(p^N) \leq \alpha
\ell(p)$. Thus $\cost_{\opt_{N}}(\sigma^N)
\leq \alpha \cdot \cost_{\opt}(\sigma)$. Summarizing:
\begin{equation*}
\E[\cost_{\algon}(\sigma) ] \leq
\E[\cost_{\algon_{N}}(\sigma^N)]
\leq r\cdot \cost_{\opt_{N}}(\sigma^N) +c
\leq \alpha r \cdot \cost_{\opt}(\sigma) +c.
\end{equation*}
\end{proof}

\begin{corollary} \label{corol:MTS1}
There is an $O(\log\Delta \log\log \Delta \cdot \log n\log\log
n)$-competitive randomized MTS algorithm for MTS defined on metric
spaces with diameter $\Delta$.
\end{corollary}
\begin{proof}
Apply Theorem~\ref{thm:path-approx} on the original metric and obtain an
$O(\log \Delta \log\log
\Delta)$ path embedding into an ultrametric of size  $\Gamma(n) = \poly(n)$.
This ultrametric has $O(\log
\Gamma(n)\, \log\log \Gamma(n))$ competitive algorithm  \cite{FM00}. Now
apply  Proposition~\ref{prop:MTS-appl} to obtain the claim.
\end{proof}

\medskip

The Group Steiner Tree Problem (GST)~\cite{RW90} can be stated as
follows: Given a graph $G=(V,E)$ on $n$ vertices with a weight
function $c:E\rightarrow \mathbb{R}_+$, and subsets of the
vertices $g_1,\ldots, g_k \subset V$ (called \emph{groups}), the
objective is to find a minimum weight subtree $T$ of $G$ that
contains at least one vertex from each $g_i$, $i\in [k]$. Under
certain standard complexity assumptions, this is hard to
approximate by a factor better than $\max\{ \log^{2-\eps} k,
\log^{2-\eps} n\}$~\cite{HK03}. The current best upper bound on
the approximation factor is $O(\log^2n \log k )$
\cite{GKR98,FRT03}. In \cite{GKR98}, an $O(\log n \log k)$
approximation algorithm for tree metrics is given, and the general
case is reduced to tree metrics using Theorem~\ref{thm:bar98}.
Again, we show that it is actually sufficient to use multi
embedding for this problem.

As a first step we observe that the problem can be easily cast in terms of
metric spaces instead of graphs: Given a graph $G=(V,E)$ with weights
$w:E\to \mathbb{R}_+$, let
$M=(V,d)$ be the shortest path metric induced by $G$ and $w$ on $V$.
A tree
$T$ in $M$ can be transformed into a tree $\hat{T}$ in $G$ such that the
total weight in $\hat{T}$ is not larger than the total weight in $T$, and
$\hat{T}$ contains all the vertices in $T$. This is done by replacing each
edge in $T$ by the shortest path between its endpoints in $G$, and taking a
spanning tree of the resulting subgraph. It therefore suffices to solve GST
on metric spaces.

\begin{proposition} \label{prop:gsp-appl}
Assume that a metric space $M$ is $\alpha$ path embedded in a metric space 
$N$.
Assume in addition that there is a [randomized] polynomial time
$r$ approximation algorithm for any GST instance with $k$ groups
defined on $N$. Then there exists a [randomized] polynomial time
$2\alpha r$-approximation algorithm for any GST instance with $k$ groups
defined on $M$.
\end{proposition}
\begin{proof}
We construct an approximation algorithm $\A$ for the instance
\linebreak
$\sigma=(M; g_1, \ldots, g_k)$ as follows. Denote by
$\f:N\to M$ the $\alpha$ path embedding of $M$ in $N$. Consider
the following instance of GST: $\sigma_N=(N;\f^{-1}(g_1), \ldots
\f^{-1}(g_k))$. Let $\A_{N}$ be an $r$-approximation algorithm for
$\sigma_N$. Let $T_N= \A_N(\sigma_N)$ be the tree constructed by
$A_N$. Denote by $\f(T_N)$ the image graph of $T_N$. \Ie, if
$T_N=(V_N,E_N)$, then $\f(T_N)=(\f(V_N), \{ \f(u) \f(v)|\; u v\in
E_N\})$. The graph $\f(T_N)$ is a connected and its weight is at
most the weight of $T_N$. It also spans at least one
representative form each group. Algorithm $\A$ returns a spanning
tree of $\f(T_N)$. This tree is a feasible solution and it
satisfies $\cost_{\A}(\sigma)\leq \cost_{\A_{N}}(\sigma_N)$.

Consider the tree $\opt(\sigma)$, double each edge in $\opt(\sigma)$ and 
take
an Euler tour $p$ of this graph. There exists a path in $N$, $p_N$, as in 
the
statement of Definition~\ref{def:path-approx}, such that $\f(p_N)=p$. The
path $p_N$ is a connected graph and spans at least one representative from
each group $\f^{-1}(g_j)$. As the weight of $p$ is twice the weight of
$\opt(\sigma)$, we have
\[\cost_{\opt_{N}}(\sigma_N) \leq  \ell(p_N) \leq
\alpha \ell(p)\leq 2\alpha \cost_{\opt}(\sigma). \]
Summarizing:
\begin{equation*}
\E[\cost_{\A}(\sigma) ]\leq \E[\cost_{\A_{N}}(\sigma_N)]
\leq r\, \cost_{\opt_{N}}(\sigma_N)  \leq 2\alpha r \,
\cost_{\opt}(\sigma) .
\end{equation*}
\end{proof}

\begin{corollary} \label{corol:gsp1}
There is an polynomial time $O(\log\Delta \log\log \Delta \log n\log k)$ 
approximation
algorithm for GST on metric spaces with diameter $\Delta$.
\end{corollary}
\begin{comment}
\begin{proof}
We apply Proposition~\ref{prop:gsp-appl} to
Corollary~\ref{cor:path-approx} and so we obtain $\alpha=\lambda$
and $\Gamma = \poly(n)$. Using the algorithm for trees from
\cite{GKR98} with $r(m)=O(\log m \log k)$, we obtain the claim.
\end{proof}
\end{comment}

\section{Multi Embedding into Ultrametrics} \label{sec:path-approx}

The following theorem is a restatement of
Theorem~\ref{thm:path-approx} in a more general form.

\begin{theorem}
Given any metric space $M=(V,d)$ on $|V|=n$ points and  diameter
$\Delta$, for any $t\in \mathbb{N}$, $M$ is $O(t\min\{\log \Delta,
\log n\})$ path embedded into an efficiently constructible
{ultrametric} of size  $\Gamma \leq n^\beta$, where \( \beta = \min \{(\log
n)^{1/t}, [t \log (4\Delta)]^{2/t}\}. \)
\end{theorem}
\begin{proof}
The construction of the multi-embedding is motivated by the
construction of subspaces approximating {ultrametric} in
\cite{BBM01,BLMN03}, but instead of deleting points, we
\emph{duplicate} them. We then prove the bounds on the
path distortion.

Let $\Delta$ be the diameter of $M$. Let $x$ and $\bar{x}$ be two points
realizing the diameter of $M$, and assume without loss of generality that
$|\{ y\in M:\ d(x,y)<\Delta/4\}|\leq n/2$ (otherwise, switch the roles of 
$x$
and $\bar{x}$). Define a series of sets $A_0=\{x\}$, and for $i\in
\{1,2,\ldots,t \}$, $A_i=\{y\in M |\ d(x,y)< i \Delta /4t\}$, and ``shells"
$S_i=A_i\setminus A_{i-1}$. Let $|V|=n$ and let $\eps_i= |A_i|/n$.

The algorithm for constructing the multi-embedding works as
follows: Choose a shell $S_{i}$, $i\in [t]$. Recursively,
construct a multi-embedding of the sub space $A_{i}$ into {an
ultrametric} $T_1$ and a multi-embedding of the subspace
$V\setminus A_{i-1}$ into {an ultrametric} $T_2$. To construct the
multi-embedding for $M$, we construct {an ultrametric} $T$ with
root labelled with $\Delta$, and two children, one is $T_1$ and
the other is $T_2$. This is a multi-embedding since the points in
$S_i$ are essentially being ``duplicated" at this stage. Note that
this is a non-contractive multi-embedding.

Next we prove an upper bound on the size of the resulting {ultrametric}
$T$, assuming that the shell was chosen carefully enough. The bound we
prove is $n^\beta$, where $\beta = \min \{(\log n)^{1/t}, [t \log
(4\Delta)]^{2/t}\}$.

We begin with the first bound. Let
$\beta=\beta(n)= (\log n)^{1/t}$. The proof proceeds by induction
on $n$ (whereas $t$ is fixed).
There must exist $i\in [t]$ such that $\eps_{i-1} \geq \eps_i^\beta$. 
Indeed,
note that $n^{-1}\leq \eps_0 \leq \eps_{t+1} \leq 1/2$. Assume for the
contrary that $\eps_{i-1}<\eps_i^\beta$ for all $i\in[t]$, then
\[ \eps_0 <\eps_1^\beta <\cdots \eps_t^{\beta^t}\leq
(\frac{1}{2})^{\log n} =\frac{1}{n} ,\]
which is a contradiction. Therefore we can fix $i$ such that $\eps_{i-1} 
\geq
\eps_i^\beta$. Inductively, assume that the recursive process results in at
most $(\eps_{i}n)^{\beta(\eps_i n)} \leq (\eps_{i}n)^\beta$
leaves in $T_1$ and at most
$((1-\eps_{i-1})n)^{\beta((1-\eps_{i-1})n)}\leq ((1-\eps_{i-1})n)^{\beta}$
leaves in $T_2$. So \( |T|\leq (\eps_{i}^{\beta}
+ (1-\eps_{i-1})^{\beta}) n^{\beta}\). Since $\eps_{i-1} \geq
\eps_{i}^\beta$, we have \( \eps_{i}^{\beta}+ (1-\eps_{i-1})^{\beta} \leq
\eps_{i-1} + (1- \eps_{i-1})=1 \) and we are done.

We next prove the second bound. Let $\beta=\beta(\Delta) = [t \log
(4\Delta)]^{2/t}$. The proof is by induction on (the rounded value of)
$\Delta$. We claim that
\begin{eqnarray}\label{eq:condition}
\exists i\in [t] \qquad \eps_{i-1}\ge \eps_{i}^{\beta(\Delta/2)}
n^{\beta(\Delta/2)- \beta(\Delta)}.
\end{eqnarray}
Indeed, assume for the contrary that no such $i$ exists. Set $a=\log
(2\Delta)\ge 1$, so that $\beta(\Delta/2)=(ta)^{2/t}$ and $\beta(\Delta) =
[t(a+1)]^{2/t}$. Denote $b=n^{(ta)^{2/t}-[t(a+1)]^{2/t}}$ and 
$c=(ta)^{2/t}$.
The opposite of \eqref{eq:condition} then becomes
$\eps_{i-1}<\eps_i^c b$, for any $i\in[t]$. Iterating this $t$ times
we get:
\[
\frac{1}{n}=\eps_0 < \eps_t^{c^t} b^{1+c+c^2+\ldots+c^{t-1}} \le
\eps_t^{c^t} b^{c^{t-1}} \le b^{c^{t-1}} .
\]
So that:
\[
n^{(ta)^{2-2/t}\left[[t(a+1)]^{2/t}-(ta)^{2/t}\right]}<n,
\]
but this is a contradiction, since an application of the mean value theorem
implies the existence of $\xi \in [a,a+1]$, for which
\begin{multline*}
(ta)^{2-2/t}\left[[t(a+1)]^{2/t}-(ta)^{2/t}\right] = % \\ =
   (ta)^{2-2/t} [2 t^{-1+2/t}\xi^{-1+2/t}]
=  2 (ta) \left(\frac{a}{\xi}\right)^{1-2/t} \ge ta \ge 1 .
\end{multline*}

Choose an index $i\in \{1,\ldots, t\}$ satisfying (\ref{eq:condition}). 
Since
$i\le t$, $\Delta(A_{i})\le \Delta/2$. The choice of the index $i$, and 
using
the inductive hypothesis, gives the required lower bound on the cardinality
of $T$ since:
{%\setlength\arraycolsep{1pt}
\begin{align*}
|T|
% &\le |A_{i}|^{\beta(\Delta(A_{i}))}+(n-|A_{i-1}|)^{\beta(\Delta(M\setminus 
% A_{i-1}))}\\
\le (\eps_{i}n)^{\beta(\Delta/2)}+[(1-\eps_{i-1})n]^{\beta(\Delta)} \le
\eps_{i-1}n^{\beta(\Delta)}+(1-\eps_{i-1})n^{\beta(\Delta)}\le
n^{\beta(\Delta)}.
\end{align*}}

We note that the running time of the algorithm above is $O(n^2)$
on each vertex in the tree and therefore $O(n^{\beta+2})$ for the
whole tree . A slight variation on this algorithm (and a more
careful analysis) has an ${O}(n^{\max\{\beta,2\}})$ running time.

\begin{comment}
To reach path-distortion of $O(n^\beta \log \log n)$,
we don't need to compute the diameter.
Instead
it suffices to take arbitrary $x$ and the furthest point $\bar{x}$ . So for 
each vertex in the HST
we only need to invest $O(n)$ time. This gives $O(n^{\beta+1})$.
To improve upon this we do a more careful inductive argument
\begin{multline*}
T(n)\leq T(\eps_i n)+ T((1-\eps_{i-1})n) + O(n) \\
   \leq c_\beta \left(\eps_{i}^{\max\{2,\beta\}} 
+(1-\eps_{i-1})^{\max\{2,\beta\}}\right) n^{\max\{2,\beta\}} +O(n) \\
  \leq \c_\beta \left(\eps_{i-1}^{\max\{2/\beta,1\}} 
+(1-\eps_{i-1})^{\max\{2,\beta\}}\right ) n^{\max\{2,\beta\}} +O(n) \\
\leq c_\beta n^{\max\{\beta,2\}}
\end{multline*}
\end{comment}

The multi-embedding described above has the following properties:
\begin{enumerate}
\item The multi-embedding is non-contractive.
\item The tree structure defining the {ultrametric} is a binary
tree.\footnote{Note that more generally, any {ultrametric} can be
defined by a binary tree.}
\end{enumerate}
Let $u$ be an internal vertex in the binary tree defining the ultrametric,
and $T$ the subtree rooted at $u$.
We can rename the subtrees rooted with the children of $u$, as $T_1$
and $T_2$ such that:
\begin{enumerate}
\setcounter{enumi}{2}
\item
\label{item:stam} Let $x$ and $y$ two points in $M$. If $\emptyset \neq 
\f^{-1}(x) \cap T \subset T_1$
and $\emptyset \neq \f^{-1}(y) \cap T \subset T_2$, then
$d(x,y)\geq \Delta(T)/4t$.
\item $|\f(T_1)| \leq |\f(T)|/2$.
\item
$\Delta(T_1)\leq \Delta(T)/2$.
\end{enumerate}

We next show, using the properties above, that the path distortion
of this multi embedding is at most $8t \log \min\{n,\Delta\}$. Let
$p=\la u_1, u_2,\ldots, u_{m} \ra$ be a path in $M$ whose length
is $\ell(p)$ . We construct a path $\bar{p}$ on the leaves of $T$
whose length satisfies $\ell(\bar{p}) \leq 8t \log
\min\{n,\Delta\} \cdot \ell(p)$. The proof proceeds by induction
on the height of the tree defining the {ultrametric}.

\begin{figure}[ht]
\centering
\input{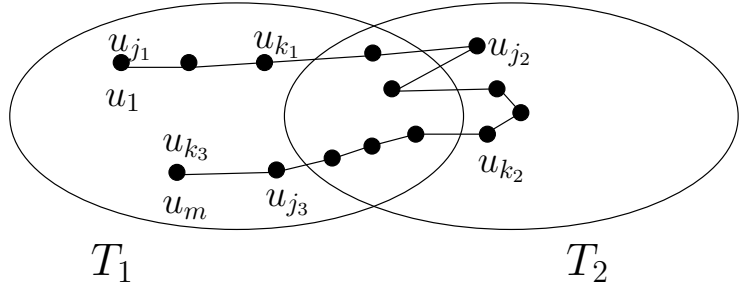}
\caption{A partition of the path to subpaths.}
\label{figure}
\end{figure}

We partition $p$ into sub-paths as follows. Define a sequence of indices and 
a
sequence of sub-trees of the root: Let $j_1=1$ and let
$\hat{T}_1\in\{T_1,T_2\}$ be the subtree of the root that includes the
longest prefix of $p$. Assume inductively that we have already defined
$j_{i-1}$ and $\hat{T}_{i-1}\in\{T_1,T_2\}$. Define $j_i$ to be the minimum 
index
such that $u_{j_i}$ is the first point in $p$ after $u_{j_{i-1}}$ with no
representative in $\hat{T}_{i-1}$. Let $\hat{T}_i\in\{T_1,T_2\}$ be the 
other
subtree of the root. Assume this process is finished with $j_{s}$,
$\hat{T}_{s}$. Next we define another sequence of indexes: $k_{s}=m$, for
$i<m$ we define $k_i$ to be the largest number, smaller than $j_{i+1}$, such
that $u_{k_i}$ does not have a representative in $\hat{T}_{i+1}$. By the
construction of $\hat{T}_i$, we have that $j_i \leq k_i$ and $u_{k_i}$ has a
representative in $\hat{T}_i$.  See Figure~\ref{figure} for example of such 
partition.
We have partitioned $p$ into sub-paths $(\la
u_{j_i},\ldots, u_{k_i} \ra)_i$ and $(\la u_{k_i},\ldots, u_{j_{i+1}}
\ra)_i$. Informally, a sub-path $\la u_{j_i},\ldots, u_{k_i} \ra$ will be
realized in $\hat{T}_i$, while sub-path $\la u_{k_i},\ldots, u_{j_{i+1}} 
\ra$
will be realized in $T_1$.

More formally,  let $L=\ell(p)$,  $L_{i_1,i_2}=\ell(\la
u_{i_1},u_{i_1+1},\ldots, u_{i_2} \ra)$, $n=|\f(T)|$, $n_1=|\f(T_1)|$,
$n_2=|\f(T_2)|$, and $\Delta=\Delta(T)$, $\Delta_1=\Delta(T_1)$,
and $\Delta_2=\Delta(T_2)$. We construct by induction on the tree structure
$T$ a path $\bar{p}$ in $T$ whose length satisfies \( \bar{L}=
\ell(\bar{p}) \leq 8t \log \min\{\Delta,n\} \cdot L . \)

By the induction hypothesis it is possible to construct for any
$i$, a path in $\hat{T}_i$ of representatives of $\la
u_{j_i},\ldots,u_{k_i} \ra$ whose length is
\[
\bar{L}_{j_i,k_i} %\leq L _{j_i,k_i} \cdot 8t \log \min \{n_i,\Delta_i\}
\leq L_{j_i,k_i} \cdot 8t  \log \min\{n,\Delta\}.
\]
Next, for any $i$, we construct a path of representatives of $ \la
u_{k_i}, \ldots ,u_{j_{i+1}}\ra$. Note that $u_{k_{i}+1},\ldots,
u_{j_{i+1}-1}$ have representatives in both $\hat{T}_i$ and
$\hat{T}_{i+1}$. Therefore, we construct inductively a path from a
representative of $u_{k_i+1}$ to a representative of
$u_{j_{i+1}-1}$ in $T_1$, so $\bar{L}_{k_i,j_{i+1}} \leq L
_{k_i,j_{i+1}} 8t \log\min\{n_1,\Delta_1\}$. We then connect the
representative of $u_{k_i}$ with the representative of $u_{k_i+1}$
and the representative of $u_{j_{i+1}-1}$ with the representative
of $u_{x_{i+1}}$, each such edge is of length at most the diameter
of $T$, $\Delta$. We have therefore constructed a path of
representatives of $ \la u_{k_i}, \ldots, u_{j_{i+1}}\ra$ whose
length is \( \bar{L}_{k_i,j_{i+1}} +2\Delta\).

Since $u_{k_i}$ does not have a representative in $\hat{T}_{i+1}$
and $u_{j_{i+1}}$ does not have representative in $\hat{T}_i$, we conclude
using
property~(\ref{item:stam}) above, that $d_M(u_{k_i},u_{j_{i+1}})
\geq \Delta /4 t$, and so $\Delta \leq 4t \cdot L_{k_i,j_{i+1}}$. To
summarize,
\begin{align*}
\bar{L}_{k_i,j_{i+1}} & \leq   L_{k_i,j_{i+1}} 8t \log
\min\{n_1,\Delta_1\}
+2\Delta \\
& \leq L_{k_i,j_{i+1}} 8t \bigl( \log \min\{n/2, \Delta/2\} + 1\bigr) \\
&= L_{k_i,j_{i+1}} 8t  \log \min\{n,\Delta\}
   .
\end{align*}
We conclude,
\begin{align*}
\bar{L} &= \sum_{i=1}^s \bar{L}_{j_i,k_i} + \sum_{i=1}^{s-1}
\bar{L}_{k_i,j_{i+1}} \\
& \leq 8t \log \min\{n,\Delta\} \cdot  \Bigl(\sum_{i=1}^s 
L_{j_i,k_i}+\sum_{i=1}^{s-1} L_{k_i,j_{i+1}} \Bigr )
\\ & =  8t \log \min\{n,\Delta\} L  .
\end{align*}
\end{proof}

We end the discussion on multi embedding into
{ultrametrics} with the following impossibility result.
\begin{proposition} \label{prop:inapprox}
Consider the metric defined by a simple $N$-point path. Then any
$\alpha$ path-embedding of this metric in an
{ultrametric} must have $\alpha =\Omega( \log n)$.
\end{proposition}
\begin{proof}
Let $M=\{v_1,v_2,\ldots,v_n\}$ be the metric space on $n$ points
such that $d_M(v_i,v_j)=|i-j|$. We prove that for any non-contractive
multi-embedding into {an ultrametric} $T$, any path of representatives
of $\la v_1, v_2,\ldots,v_n \ra$ is of length at least
$g(n)=\tfrac{n}{2} \log n$.

The proof proceeds by induction on $n$. For $n=1$ the claim is trivial. For
$n>1$, let $\la v'_1,v'_2,\ldots, v'_n \ra$ be a path of representatives in 
$T$.
Let $u=\lca_T(v'_1,v'_n)$, $\Delta(u)=d_T(v'_1,v'_n)\geq
d_M(v_1,v_n)=n-1$. Let $T_1$ be the subtree of the child of $u$ that 
contains
$v'_1$. $T_1$ does not contains $v'_n$. Let $i_1<n$ be the maximal $i$ such
that $\{v'_1,\ldots,v'_{i_1}\}\subset T_1$. As
$i_1+1$ is not contained in $T_1$, it must be that
$d_T(v'_{i_1},v'_{i_1+1})\geq \Delta(u)\geq n-1$. By the induction 
hypothesis
\begin{align*}
\ell_T(\la v'_1,\ldots, v'_{i_1} \ra) &\geq g(i_1), & \ell_T(\la
v'_{i_1+1},\ldots, v'_{n} \ra) &\geq g(n-i_1).
\end{align*}
Since $g$ is a convex function, $(g(i_1)+g(n-i_1))/2 \geq g( (i_1
+(n-i_1))/2)= g(n/2)$. We conclude
\begin{multline*}
\ell_T(\la v'_1,\ldots, v'_{n} \ra)  \geq g(i_1)+ (n-1)+ g(n-i_1)
\\     \geq 2 g(n/2) + n-1
     = 2 \tfrac{n}{4} \log \tfrac{n}{2} +(n-1)
     \geq \tfrac{n}{2} \log n .
\end{multline*}
\end{proof}

\section{Multi-Embedding into Trees} \label{sec:trees}

In this section we consider multi embeddings into \emph{arbitrary
tree metrics}. We only have preliminary results. Specifically, we
only consider two important types of metric spaces: expander
graphs  and the discrete cube with the Hamming metric, for which
we obtain better results. For both of them the preceding sections
proved an upper bound of $O(\log\log n \log\log\log n)$ and a
lower bound of $\Omega(\log \log n)$ on the path distortion of 
multi-embeddings
into {ultrametrics}  (the lower bound follows since both metrics contain
a path of length $\Omega(\log  n)$).

We begin with the observation that for $\Gamma = \infty$ it is
easy to obtain $1$ path embedding of any finite metric space into
trees. This is achieved by defining an infinite tree
metric as follows: joining all possible finite paths with a common
root, where the first node in the path is connected with an edge
weight of $\Delta/2$ to the root. Moreover,

\begin{proposition} \label{prop:trees}
Given a metric space $M$ defined by an unweighted graph of maximum
degree $d$ and diameter $\Delta$, and let $s\in\mathbb{N}$. Then $M$ can be
$(2+\frac{\Delta}{s})$ path-embedded into a tree metric of size
$n d^s$.
\end{proposition}
\begin{proof}
Along the lines of the construction described above, we take all paths of 
length
$s$, and join these with a common root, where the first node in the path is
connected with an edge weight of $\Delta/2$ to the root. Obviously, there 
are
at most $n d^s$ such paths.
Notice
that our choice  of weights to the edges adjacent to the root guarantees
that distances in
the resulting tree are no smaller than the original distances.
We next claim that the path distortion is at most
$(2+\frac{\Delta}{s})$. To see this,
consider a path
$p=\langle v_1, \ldots v_\ell \rangle$ of length $\ell$. We partition $p$
into sub-paths of length $s$: $p=p_1p_2\cdots p_t$, where $t=\lceil \ell/s
\rceil$, $p_j=\langle v_{(j-1)s+1}, \ldots, v_{js} \rangle$ for $j<t$, and
$p_t=\langle v_{(t-1)s+1},\ldots, v_\ell \rangle$. Now the sub-path $p_j$ is
mapped to the appropriate path in the tree. Note that the length of the 
image
path is $2\ell+  (t-1)\Delta$.
\end{proof}

This simple fact is particularly interesting for its implication
for expander graphs. Let $G$ be an $(n,d,\gamma d)$ graph, i.e., a
$d$-regular, $n$-vertex graph whose second eigenvalue in absolute
value is at most $\gamma d$. It is known \cite{Ch89} that such a
graph  has diameter at most $1+\log_{1/\gamma}n$, and so we obtain:
\begin{corollary}
Any $(n,d,\gamma d)$-graph has $3$ path embedding in a tree of size
$d n^{1+\log_{1/\gamma}d}$.
\end{corollary}

We also note that for the trees constructed in the proof of
Proposition~\ref{prop:trees}, it is particularly easy to obtain
a better approximation algorithm  for GST.
\begin{lemma} \label{lem:approx-group}
Consider a tree metric $M=(V,d)$, where $V=P_1\cup P_2\cup \cdots \cup
P_\ell$,  $P_i=\langle v^i_1,v^i_2,\ldots, v_s^i\rangle$ is an unweighted
simple path of length $s$, and $d(v^i_1,v^j_1)=\Delta$ for $i\neq j$. Then 
an
instance of the GST with $k$ groups defined on $M$ has
$(1 + \frac{2s}{\Delta})(1+ \ln k)$ approximation algorithm.
\end{lemma}
\begin{proof}
Consider a GST instance $g_1,\ldots g_k$ defined on
$M$. We first check whether there is a solution that is completely
contained in one $P_i$. This can be checked in polynomial time by
noting that if an optimal solution is contained in one $P_i$ then
it is an interval. Thus all is needed to be checked are  $\ell
\binom{s+1}{2}$ intervals.

Otherwise, the optimal solution intersects $t>1$ of the paths $P_1,\ldots
P_\ell$. Define a Hitting Set instance whose ground set is
$\{P_1,\ldots,P_\ell\}$ and the subsets are $g'_1,\ldots, g'_k$, where
$g'_i=\{P_j;\; P_j\cap g_i \neq \emptyset \}$. It follows that the optimal
cost of the hitting set problem is at least $t-1$.
The Hitting Set problem has a polynomial time $1+\ln
k$ approximation algorithm~\cite{Jo74,L75}. Let $S$ be the approximate
solution for the hitting set. We define a solution for the GST
instance
by taking a natural path over $\cup \{P_i;\; P_i\in
S\}$. Note that its length is at most $(\Delta+2s) |S| \le
(\Delta+2s)(t-1)(1+\ln k)$. But the cost of the optimal GST
algorithm is at least $(t-1)\Delta$.
\end{proof}

\begin{corollary}
For fixed $d>\lambda$,
There exist constants $c=c_{d,\lambda}$, $C=C_{d,\lambda}$, and polynomial
$p(t)=p_{d,\lambda}(t)$ such that GST on
$(n,d,\lambda)$ graphs has $p(n)$-time $(C\log k)$ approximation algorithm, 
and it
is NP-hard to approximate within a factor of $c\log k$.
\end{corollary}
\begin{proof}
The approximation algorithm follows from Proposition~\ref{prop:trees} and
Lemma~\ref{lem:approx-group}, by setting $s=\Delta$.
The hardness result follows since an $(n,d,\lambda)$
graph contains a subset of $n^{\Omega_{d,\lambda}(1)}$ points that is 
$O_{d,\lambda}(1)$
approximated by an equilateral space~\cite{BLMN03}.
GST on equilateral space is equivalent to a standard Hitting Set problem,
which is NP-hard to
approximate within a factor of $c\ln k$ \cite{RS97,LY94}.
Usage of points not in this
subspace (``Steiner points") can improve the approximation factor by at most
a factor of two~\cite{RW90,G01}.
\end{proof}

\medskip

We next examine multi-embedding of the $h$-dimensional hypercube with 
$n=2^h$
vertices. Using Proposition~\ref{prop:trees} with $s=h/\log h$, and using $d
= h$ and $\Delta = h$, we obtain $(\log \log n+2)$ path embedding into trees
of size
$\Gamma(n)\leq n^2$. Using Lemma~\ref{lem:approx-group} it also implies a
polynomial time $O(\log k \log \log n)$ approximation algorithm to GST
on the cube. Similarly to the expander graphs, it is
hard to approximate instances of GST on the cube
to within a $c\log k$ factor, for some constant $c>0$, since the cube
contains a subset of $n^{\Omega(1)}$ points that is $O(1)$ approximated by 
an
equilateral space.

\section{Discussion} \label{sec:conclusion}

An interesting open problem is to determine worst case bounds for
path distortion of multi embedding into trees of polynomial size.
As indicated by the case of expander graphs, such bounds may be
better than those for {ultrametrics}. 

Results on multi-embedding into trees directly reflect on the
approximability of GST. As shown for expanders and hypercubes, it
is possible that for special classes of metric spaces, a
combination of improved path embedding and a specialized solution
would yield (nearly) tight upper bounds. Our approach to show the
(near) tightness of the results in those cases stems from metric
Ramsey-type considerations (i.e. the existence of large
approximately equilateral subspace). Such considerations are in
fact more general and may lead to more results of this flavor.
\footnote{In \cite{BLMN03} it is shown that any metric space
contains a ``large" subspace which is approximately an
ultrametric, or a $k$-HST. Such trees were used in \cite{HK03} to
prove inapproximability results for GST. It is plausible that
these techniques can be combined to obtain tight bounds for GST in
specific metric spaces.} 

Multi-embedding into {ultrametrics}, also implies multi embedding into
$\ell^d_p$, where $d=O(\log n)$ with similar
path distortion~\cite{blmn-lowdim}. It is natural to ask whether better
path distortion is possible for multi-embedding into $\ell_1$ or $\ell_2$.
Further study of multi-embeddings in other settings and their applications
seems an attractive direction for future research.

\subsubsection*{Acknowledgments} We thank Robi Krau\-th\-gamer
and Assaf Naor for fruitful discussions.

\bibliographystyle{siam}
\bibliography{path}
\end{document}